# Evolution of wartime discourse on Telegram: A comparative study of Ukrainian and Russian policymakers' communication before and after Russia's full-scale invasion of Ukraine

*Mykola Makhortykh, University of Bern, mykola.makhortykh@unibe.ch*

*Aytalina Kulichkina, University of Vienna, aytalina.kulichkina@univie.ac.at*

*Kateryna Maikovska, University of Vienna, kateryna.maikovska@univie.ac.at*

**Abstract:** This study examines elite-driven political communication on Telegram during the ongoing Russo-Ukrainian war, the first large-scale European war in the social media era. Using a unique dataset of Telegram public posts from Ukrainian and Russian policymakers (2019-2024), we analyze changes in communication volume, thematic content, and actor engagement following Russia's 2022 full-scale invasion. Our findings show a sharp increase in Telegram activity after the invasion, particularly among ruling-party policymakers. Ukrainian policymakers initially focused on war-related topics, but this emphasis declined over time In contrast, Russian policymakers largely avoided war-related discussions, instead emphasizing unrelated topics, such as Western crises, to distract public attention. We also identify differences in communication strategies between large and small parties, as well as individual policymakers. Our findings shed light on how policymakers adapt to wartime communication challenges and offer critical insights into the dynamics of online political discourse during times of war.

**Keywords**: Telegram, war, Ukraine, Russia, policymaker, political communication

**Acknowledgements**: The paper draws on the lists of policymakers compiled by the authors as part of the "Telegram around the globe" project, led by Annett Heft, Emese Domahidi, and Pablo Jost.

**Introduction**

Over the past two decades, social media and messengers have become essential platforms for political communication (Severin-Nielsen, 2023), both among elites and the general public. These platforms have enabled new opportunities for political deliberation in digital public spheres (Schäfer, 2014), but they have also facilitated the diffusion of misleading information and hate speech, which damage healthy democratic debate (McKay & Tenove, 2020). These contrasting functions of social media, as spaces for democratic deliberation and as channels for disinformation and polarization, become even more pronounced during times of war.

In wartime, political elites (e.g., policymakers) often turn to social media and messengers to mobilize support domestically and internationally, manipulate opinions, and crowd out critical views (Zeitzoff, 2017). However, the use of these platforms can vary greatly across contexts. Most recent wars involving countries with high internet penetration have been relatively short-term (e.g., the Russo-Georgian war) or constituted low-intensity counterinsurgency operations (e.g., the US-led invasion of Afghanistan). In contrast, more intense and prolonged wars have primarily occurred in the Global South (e.g., in Libya or Syria). As a result, there is limited empirical evidence on how political communication unfolds during wartime in socially mediated environments. Given the growing risk of wars returning to the Global North, there is an urgent need for such evidence to strengthen the resilience of democracies that are vulnerable to both informational and physical attacks, and to understand how autocracies, historically more prone to initiating wars (Tangerås, 2009), manage political communication during mass violence.

In this study, we address this gap by examining elite-driven political communication on a hybrid social media/messenger platform, Telegram, in Ukraine and Russia. These two countries, with high levels of internet penetration, have been engaged in a war that has been ongoing since 2014 and intensified in 2022, following Russia's full-scale invasion of Ukraine. It is the first war on such a scale in Europe since the emergence of social media, which presents a critical context for understanding how platforms are adopted and used by political elites during today's wars. To this end, we draw on a unique dataset of public posts from Telegram—the central war-related information platform in both countries (Bawa et al., 2025)—coming from all major legislative and executive actors in Ukraine and Russia from 2019 to 2024. Using descriptive statistics and computational text analysis, we examine how the volume and content of political communication on Telegram, as well as individual actor engagement, changed following the full-scale invasion and evolved throughout the course of the war.

The rest of the article is organized as follows: First, we provide a concise overview of the related work on political communication during wartime and introduce our research questions. Then, we describe our case platform, Telegram, together with the methodology we used for data collection and analysis. After it, we present our findings regarding the volume of political communication before and after the invasion, the thematic changes of the content of such communication, and the role of specific political actors in it. We conclude the article with a discussion of what our findings imply for political communication research in online environments, together with the limitations of the current study and directions for future research.

**Political communication in times of war**

Extant research on political communication during wartime has primarily focused on historical conflicts, such as the World Wars (e.g., Lasswell, 1927), or mostly pre-Web.2.0 wars in the former Yugoslavia (e.g., Kolstø, 2016) and Chechnya (e.g., Thomas, 2000). These studies, while valuable, examine periods when the information ecosystem was fundamentally different from the current landscape shaped by messengers and social media. In the past, state-controlled broadcast media played a dominant role in spreading political information, with the public's access to alternative viewpoints and real-time events being far more limited. However, the shift from centralized channels of information and propaganda to multifaceted and interactive digital media has transformed how political elites produce war narratives and how these narratives are experienced by individuals within and outside conflict zones (Ford & Hoskins, 2022).

The importance of this transformation is highlighted by a dynamically growing body of research on the impact of digital platforms on wartime political communication (Hoskins & O'Loughlin, 2010; Patrikarakos, 2017; Ford & Hoskins, 2022; Kobilke et al., 2023; Zasanska & Ivanenko, 2025). In this article, we engage with a small part of this extensive scholarship to focus on the use of social media and messengers by political elites (e.g., policymakers and political institutions). Such an emphasis is important due to the limited attention of the current research to the elite-driven online communication during modern wars, despite the known importance of elites' opinions and priorities for shaping media coverage of mass violence (Aday, 2017).

To date, most studies on the topic have looked at a few selected cases dealing primarily with short-term periods of intense violence, such as the US-led invasions in Iraq and Afghanistan (Bahar, 2020; Zappettini & Rezazadah, 2024) or the wars involving Israel in the Middle East (Manor & Crilley, 2018; Wolfsfeld, 2018; Massa & Anzera, 2023). A few recent studies (Manor, 2023, 2025; Pavliuc, 2025) have also looked at Russia's war in Ukraine—the focus of this article—but they primarily concentrated on digital public diplomacy, often examining the actions of a few selected Ukrainian actors (e.g., Volodymyr Zelensky).

In terms of methodological approaches, the extant work on the elite-driven online communication tends to rely on qualitative approaches (for an exception, see Pavliuc, 2025).

These approaches range from interview-based assessments (Wolfsfeld, 2018) to textual and visual framing analysis (Manor & Crilley, 2018) and critical discourse analysis (Zappettini & Rezazadah, 2024). Thematically, these studies often look at platforms' potential to amplify propaganda and hate speech (Wolfsfeld, 2018; Bahar, 2020) and enable the crowdfunding of wars (Manor, 2025). While these studies provide rich contextualized insights, their reliance on small-scale data results in an inability to capture patterns at scale and compare communication dynamics. Computational approaches are therefore essential to complement this body of work and enable systematic analyses of socially mediated political communication during wartime.

While research on elite-driven wartime communication is growing, several research gaps remain to be addressed. First, as noted earlier, most existing studies focus on short-term conflicts. To much degree, it is due to more intense and long-term wars occurring in regions with low internet penetration and restrictive media regimes, leading to significant differences and constrains in online communication (Lynch et al., 2014; Saleh et al., 2020). The immediate consequence of skewness is the limited understanding of how "war feeds" (Hoskins & Shchelin, 2023) would interact with elite-driven communication during a long-term war that is happening in a hyperconnected information ecosystem similar to the ones in the Global North.

Second, the focus on qualitative approaches makes it difficult to systematically assess how online platforms influence wartime political communication. Such an assessment would require processing large volumes of data, as it was done by Pavliuc (2025), who applied structural topic modelling to examine gender-based differences in communication by Ukrainian politicians on X. Most existing studies, however, look at communication by individual actors, be it politicians or institutions (e.g., the Afghan Ministry of Defense; Bahar, 2020). Another implication of the qualitative focus is that research concentrates on relatively few communication topics, usually the ones explicitly related to mass violence (e.g., military operations; Bahar, 2020). However, it results in little knowledge about the impact of platforms on communication regarding other topics that are relevant for policymaking (e.g., economy or education).

Finally, just like the mass violence itself, wartime political communication is a volatile process that results in frequent changes in elites' online discourse. For instance, Zappettini and Rezazadah (2024) showcased the evolution of different discursive strategies of the US administration on X regarding the withdrawal from Afghanistan. Similarly, Manor and Crilley (2018) in their analysis of Twitter-based communication of the Israeli Ministry of Foreign Affairs during the 2014 Gaza war demonstrated the rapid change of frames prioritised by the Ministry as the war developed. However, the degree to which political communication changes over time among different actors, especially for the long-term high-intensity conflicts, is yet to be studied.

To address these gaps, we look at online political communication during Russia's war in Ukraine, which is the first long-term and high-intensity war in the region with high internet penetration since the emergence of Web 2.0. Using a large set of public data from Telegram, we examine how political communication in Ukraine and Russia, the two countries central to the war, changes over time in terms of three key aspects: the volume of communication, its content, and the involvement of specific actors. While doing so, we explore all content coming from policymakers, without focusing just on a small range of violence-related topics. Specifically, we aim to answer the following three research questions:

*RQ1*: How has political communication activity by policymakers in Ukraine and Russia changed following Russia's full-scale invasion of Ukraine?

*RQ2*: How has the content of political discourse of Ukrainian and Russian policymakers evolved before and after the invasion?

*RQ3*: Which Ukrainian and Russian policymakers have driven key elements of the political discourse before and after the invasion?

**Case platform: Telegram**

As a case study to address our research questions, we chose Telegram. Launched in 2013 by Nikolai and Pavel Durov, two Russia-born technology entrepreneurs, Telegram combines features of a messenger and social media. It allows users to create accounts connected to their phone numbers and then participate in one-on-one conversations, group chats, or broadcast-style channels with millions of subscribers. The platform is known for its emphasis on privacy, including optional end-to-end encryption, limited content moderation, and reluctant cooperation with law enforcement agencies (Frischlich et al., 2022), even amid allegations about Pavel Durov cooperating with the Russian security services (Anin and Kondratyev, 2025).

Due to its features, Telegram was long considered a marginal platform used by niche groups, ranging from far-right communities to professional criminals. These considerations shaped the focus of the early scholarship on Telegram (Urman & Katz, 2022; Schulze et al., 2022; Roy et al., 2024). However, over time, its adoption broadened, and Telegram transitioned into a more mainstream platform. By 2025, it had reached approximately 1 billion monthly active users (Backlinko, 2024), solidifying its status as a key communication platform.

While Telegram's user base is rapidly expanding in the Global North, its initial adoption was more pronounced outside of it, including in Ukraine and Russia. While the platform had already established a significant presence in the two countries prior to 2022, its usage surged dramatically following the full-scale invasion. Telegram's perceived security and accessibility rendered it indispensable for both civilian and military coordination (Nazaruk, 2022; Ptaszek et al., 2023; Canevez et al., 2024). It has been used for many purposes: from inconspicuous war witnessing (Bareikytė & Makhortykh, 2024) and cyber resistance (Canevez et al., 2024) to Russian propaganda (Kiforchuk, 2023) and hate speech (Makhortykh & González-Aguilar, 2023).

Although a few existing studies (Yuskiv et al., 2022; Ptaszek et al., 2024; Oleinik, 2024) examined political narratives on Ukrainian and Russian Telegram channels, most of them focused on a small set of influencers. While such research provides valuable insights into wartime communication strategies, there remains no systematic, large-scale investigation of elite-driven political communication on Telegram in the context of war. By addressing the research questions outlined earlier, we aim to fill this gap by empirically examining how top-down communication from policymakers to their audiences evolved on Telegram before and after 2022.

**Method**

***Data collection and preprocessing***

To collect data, we utilized the lists of Ukrainian and Russian policymakers on the national level established by the authors as part of the *Telegram Around the Globe* project. We included individual and collective actors dealing with legislative and executive power, such as parliamentarians, national political parties, ministries, and members of national governments (more information about the political system in the two countries is provided in Appendix 2). To keep the collection feasible, we omitted policymakers on the regional level, assuming that they would likely have a lesser impact on political communication. For each policymaker, we searched for their Telegram channel. For Russia, 456 out of 683 policymakers had a channel; for Ukraine, it was 126 out of 449.

In December 2024, we used the Python *telethon* library (Lonami, 2025) to collect data from the moment each Telegram channel was established, resulting in 537,342 Telegram posts for Ukrainian policymakers and 1,450,158 for Russian ones. To make the comparison between the two countries more meaningful, we filtered out content published before 2019. It constituted a rather small part of the respective datasets: 3,841 (0.71%; Ukraine) and 7,666 (0.53%; Russia) posts. Our decision is based on 2019 being the year when the Zelensky administration and Zelensky-affiliated party, the Servant of the People, came to power with its strong focus on digital communication. Many of these newcomers were not present on Telegram before 2019, but started using it to communicate with the voters.

During data preprocessing, we kept only Telegram posts with textual content and excluded posts consisting of links to external media (e.g., videos). While such posts constituted a relatively large share of the datasets—637,424 (44%; Russia) and 160,338 (30,1%; Ukraine) posts, this decision was guided by our focus on computational text analysis. Because large-scale multimodal content analysis would require different methods and theoretical frameworks, we decided to focus on textual content to maintain a clear focus and methodological consistency. However, we consider a follow-up study looking at external content referenced in Telegram posts.

To prepare data for the analysis, we first run language detection on collected Telegram posts using the *langdetect* Python library (Nakatani, 2010). We opted for this step considering that some Ukrainian policymakers communicate online in Russian, and we wanted to explore how such uses were affected by the invasion. Similarly, we expected some policymakers in both countries to occasionally post in English to target international audiences and, especially in the case of Russia, potentially use regional languages to communicate with the voters.

Based on the language detection results (for more information, see the first subsection of Findings), we did additional preprocessing to prepare data for the computational text analysis. For the Russian dataset, we kept only Telegram posts in Russian due to the insignificant amount of content in other languages. The Ukrainian dataset was split into two subsets containing posts in Ukrainian (n = 303,090) and Russian (n = 59,673) to assess whether there were major differences in political communication depending on the language.

Finally, we used *spaCy* (Honnibal et al., 2020) lemmatizers for Ukrainian and Russian to lemmatize our data. While lemmatization is not always advantageous for computational text analysis, especially for low-inflected languages such as English (Schofield & Mimno, 2016), it is essential for Ukrainian and Russian due to their highly inflected nature. In both languages, the same words can take multiple forms when expressing grammatical relations, which limits the effectiveness of computational methods (May et al., 2016; Afanasev et al., 2025). To prevent it, we used lemmatization to resolve each word to its lemma and decrease the amount of noise.

### *Data analysis*

To answer the research questions, we combined descriptive statistics and computational text analysis. For RQ1, we applied descriptive statistics to examine how the number of Telegram posts changed over time, how many policymakers were involved in political communication, and in which languages they posted. As noted before, especially for Ukraine, with its large Russian-speaking population, the choice of language is an important indicator of party ideology and individual political position, particularly following the full-scale invasion.

To answer RQ2, we used dynamic topic modelling to examine the evolution of policymakers' narratives before and after the invasion. Dynamic topic modeling is used to analyze how topics evolve over time within a collection of documents (Blei & Lafferty, 2006). We used its implementation based on *BERTopic*, a technique introduced by Grootendorst (2022) that combines transformers and c-TF-IDF to improve document clustering. BERTopic is often used for computational text analysis (Urman & Makhortykh, 2025) due to its substantive improvement over earlier topic modelling techniques (e.g., LDA) and performance comparative for more resource-consuming large language model-based approaches (Azher et al., 2024). The full description of the BERTopic implementation is provided in Appendix 1. The topics were manually validated and interpreted by the authors. Finally, to address RQ3, we then examined which of these narratives were driven by specific parties and individuals.

**Findings**

### *Volume of political communication on Telegram in Ukraine and Russia*

Answering RQ1, we started our analysis by examining the volume of political communication on Telegram and how it changed after the full-scale Russian invasion in February 2022. Each of the two periods—pre-invasion and post-invasion—spans approximately three years of Telegram data, allowing for a meaningful comparison. Table 1 shows that in Ukraine and Russia, the volume of communication increased substantially following the full-scale invasion, with the number of posts rising more than threefold in Ukraine and more than sevenfold in Russia. This increase occurred in the overall number of posts and the median number of posts per policymaker, indicating that the heightened activity was not due to a few anomalously active actors but reflects a general tendency towards more active communication on Telegram.

**Table 1**. Descriptive statistics regarding the volume of Telegram activity among Ukrainian and Russian policymakers before and after the full-scale invasion.

| | Number of posts (in all languages) | | Number of policymakers posting | | Median number of posts per policymaker | |
|---|---|---|---|---|---|---|
| | UKR | RU | UKR | RU | UKR | RU |
| Before invasion | 89,100 | 92,849 | 85 | 137 | 401 | 200 |
| After invasion | 284,063 | 712,219 | 111 | 452 | 1,218 | 922 |

The increase in the number of posts after February 2022 was accompanied by a substantive increase in the number of policymakers usingTelegram (more than three times for Russia[1] and less than two times for Ukraine). Figures 1-3 show the distribution of Telegram accounts across specific political parties, including those of ministries and members of the government (merged under the "Government" category). For Russia (Figure 1), the prominent increase in the number of policymakers present on Telegram was observed for the ruling United Russia party (for a brief overview of each party in Ukraine and Russia, see Appendix 2). While other parties also experienced an uptick in Telegram adoption following the full-scale invasion, these increases were relatively modest. Across all Russian parties, the number of policymakers on Telegram remained rather stable, with the exception of United Russia, which saw more policymakers joining Telegram in the years after the invasion.

---

[1] When interpreting the increase for Russia, it is important to keep in mind that our dataset focuses on policymakers elected in the 2021 Russian federal elections. While many of them were re-elected (and, thus, have used Telegram for political communication before 2021), some of them were not active on the platform before being elected.

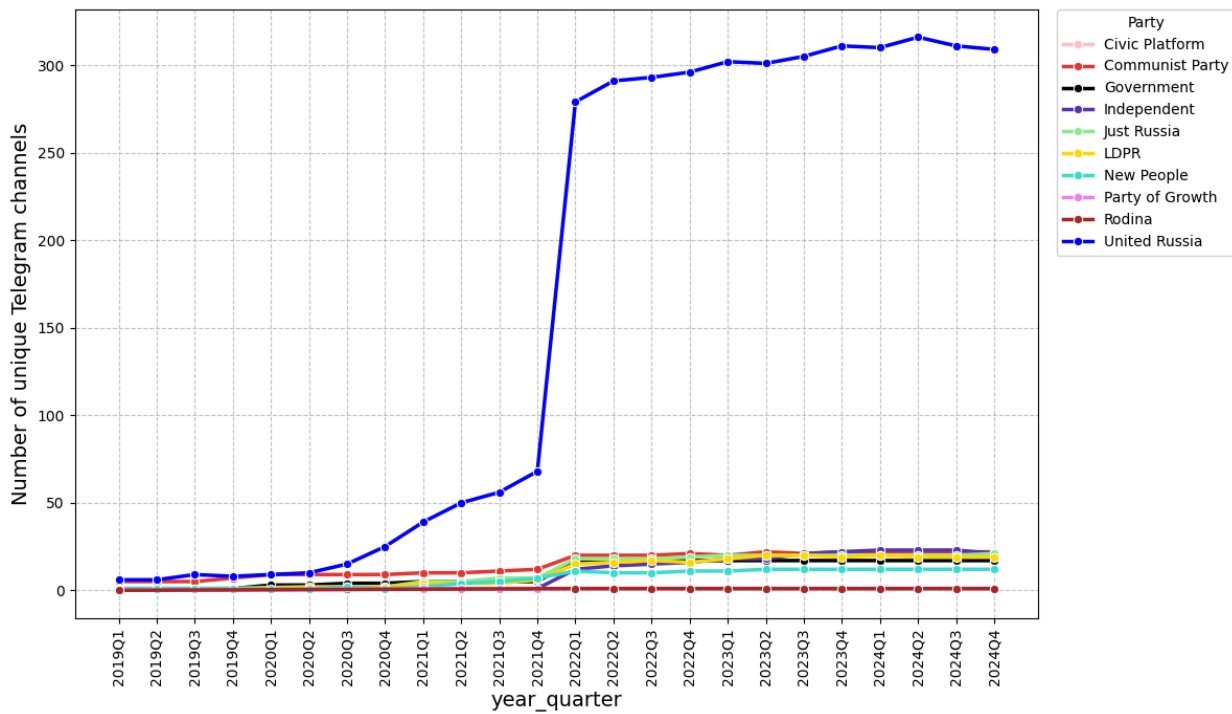

*Figure 1.* The number of unique Telegram channels among Russian policymakers posting in Russian per party per quarter.

For Ukrainian policymakers, the uptick in the number of accounts after the invasion was less pronounced due to an already present growth trend (Figures 2-3), especially for the government-related actors, the Servant of the People, and independent policymakers posting in Ukrainian (Figure 2). Following the invasion, policymakers from opposition parties, such as European Solidarity and Holos, also started using Telegram for communication, although for Holos, the activity dropped at the end of 2023. By contrast, the presumably more Russia-sympathetic opposition party Opozitsiina Platform started to be more active on Telegram only in 2023, a year after the beginning of the invasion.

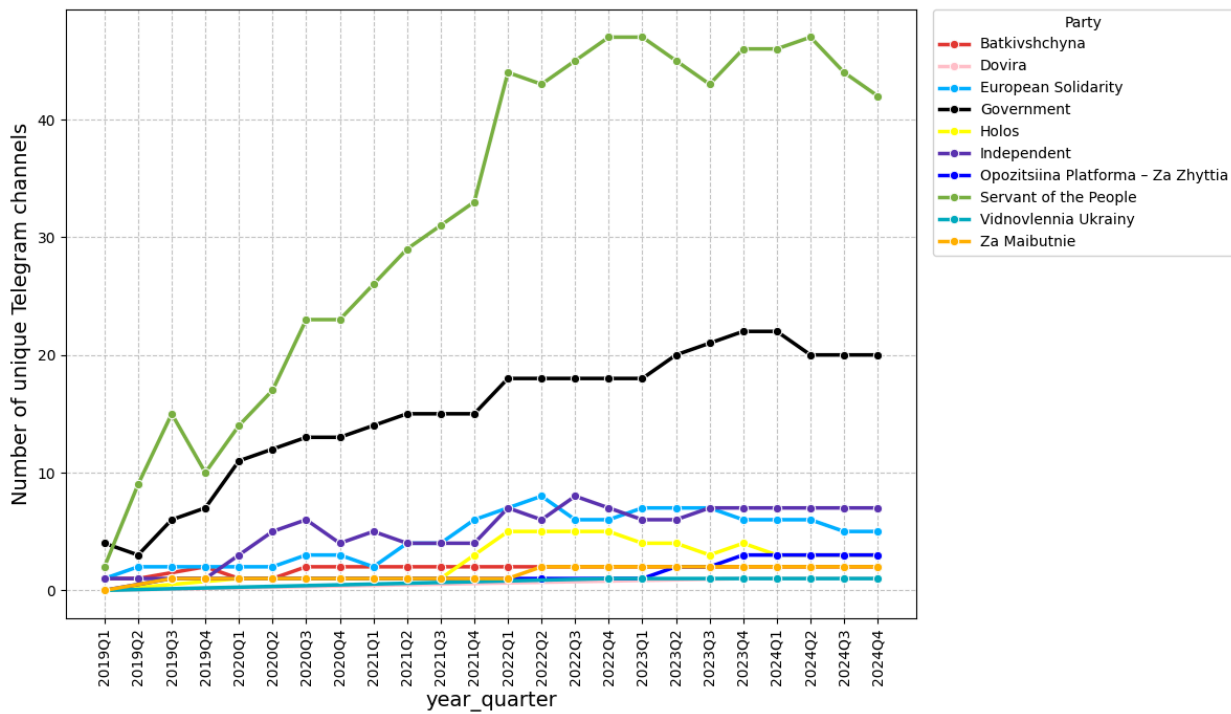

***Figure 2.*** *The number of unique Telegram channels among Ukrainian policymakers posting in Ukrainian per party per quarter.*

We observed a rather distinct dynamic among Ukrainian policymakers who posted in Russian. Before the invasion, the use of Russian gradually increased, particularly among the Servant of the People party, government officials, and independent policymakers. However, after the surge around the invasion, the number of policymakers posting in Russian has decreased for these three groups (same for Holos-related policymakers, some of whom started posting in Russian at the beginning of 2022). With some occasional exceptions, this dynamic indicates that these policymakers either stopped posting or shifted to using Ukrainian.

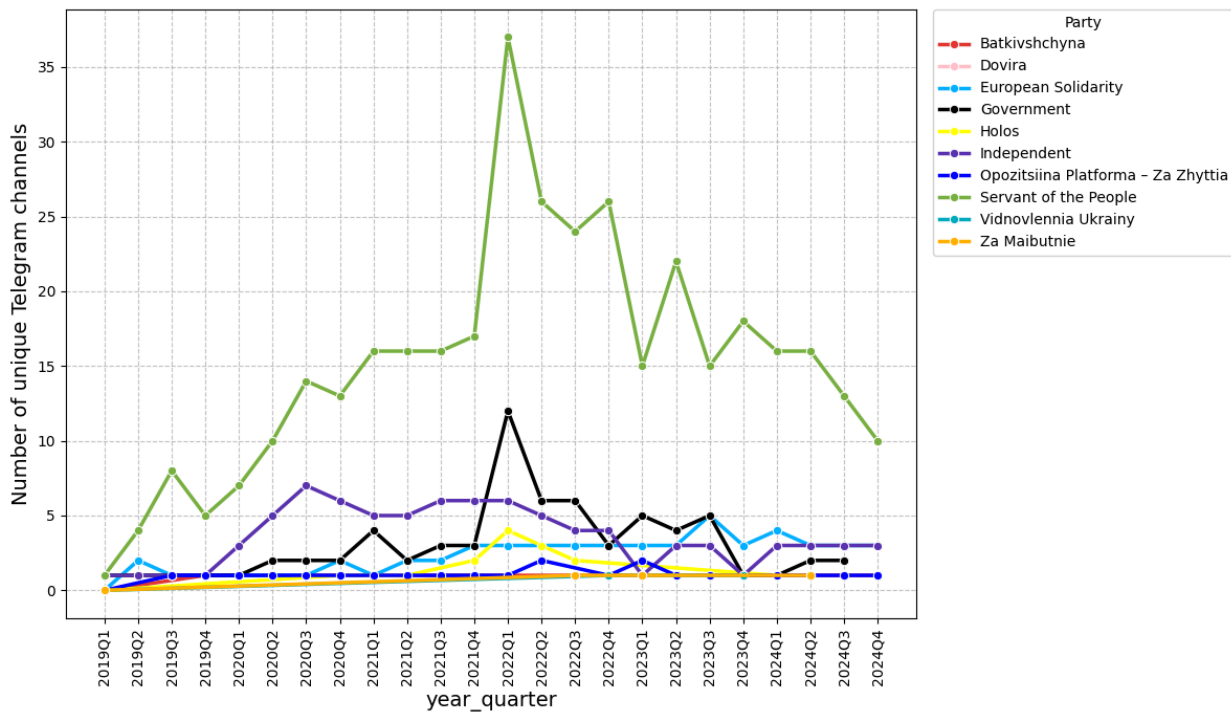

**Figure 3.** *The number of unique Telegram channels among Ukrainian policymakers posting in Russian per party per quarter.*

Finally, we looked at the volume of Telegram content communicated in specific languages (Figure 4). For Ukraine, the Russophone content started to decline already in 2020, whereas communication in Ukrainian steadily increased, reaching its peak (in absolute numbers) in 2022. After 2022, the intensity of communication among Ukrainian policymakers gradually decreased, with a slight increase in Russophone content observed in 2024. While such an increase was relatively small and the total number remains below the 2021 level, it aligns with the noted tendency regarding the decreasing support towards the predominance of the Ukrainian language in Ukraine that reached its maximum at the beginning of the war (Kulyk, 2025).

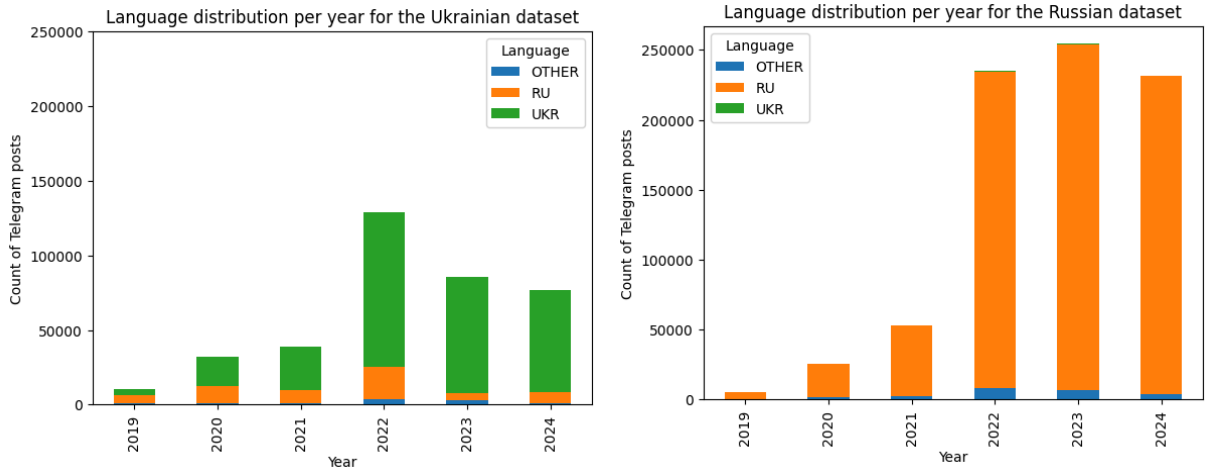

***Figure 4***. *Annual number of posts by language in the Ukrainian (left) and Russian (right) samples.*

In the case of Russia, there was close to no content in Ukrainian (Figure 4), indicating a lack of interest among Russian policymakers in addressing Ukrainian speakers. It aligns with the Kremlin narrative that portrays the Ukrainian nation and language as artificial constructs (Bohomolov & Lytvynenko, 2012). We also observed scarce communication in other languages, despite the presence of co-official languages in the Russian republics and autonomous districts represented by the policymakers; this scarcity, however, aligns with linguistic assimilation policies promoted by the Kremlin in recent years (Arutyunova & Zamyatin, 2021). In terms of the absolute number of Telegram posts, the pre-invasion period showed slow growth similar to Ukraine, while the post-invasion period was marked by a rapid and sustained increase.

***Content of political communication on Telegram in Ukraine and Russia***

To answer RQ2, we examined the distribution of content regarding specific communication topics on Telegram and its changes over time. Figure 5 demonstrates that in Russia, both before and after the invasion, most policymakers' communication was unrelated to the war. Instead, it consisted of unrelated news (e.g., about domestic politics) and discussions of the (usually positive aspects of) Russian economy and finances, volunteering initiatives, and social care programmes. This focus on positive domestic developments contrasts with frequent references to crises outside Russia, such as the failures of the US-led coalition in Afghanistan and the Kosovo war. These references were often accompanied by an emphasis on Western hypocrisy.

The news related to the war, such as frontline developments and updates on Ukrainian drone attacks, constituted a small portion of political communication. Its presence further declined in winter 2022-2023, following the liberation of Kherson and large chunks of the Kharkiv region by Ukraine. After the unsuccessful Ukrainian counteroffensive in the summer of 2023, discussions of war-related developments by Russian policymakers intensified once again. However, this topic still remained less prominent than civil topics, only marginally more common than posts wishing the audience well during national and regional holidays.

Two other topics that remained rather underrepresented in the communication of Russian policymakers were news about law enforcement and explicit references to national ideology and traditional values. Before 2022, law enforcement news was among the most common topics, highlighting its relevance for the Russian political discourse, as achievements in crime reduction are often seen as key indicators of a successful policymaker. However, after the invasion, it became much less present, which can be attributed to the topic becoming a more sensitive issue with enforcement-related actors gaining more power, and it becoming a more divisive subject compared to issues like social care. As for national ideology and traditional values, their limited presence can be attributed to relatively few posts that praise patriarchal or religious values directly. Instead, these issues were usually woven into discussions contrasting Russia with the West as part of the topic about crises outside of Russia.

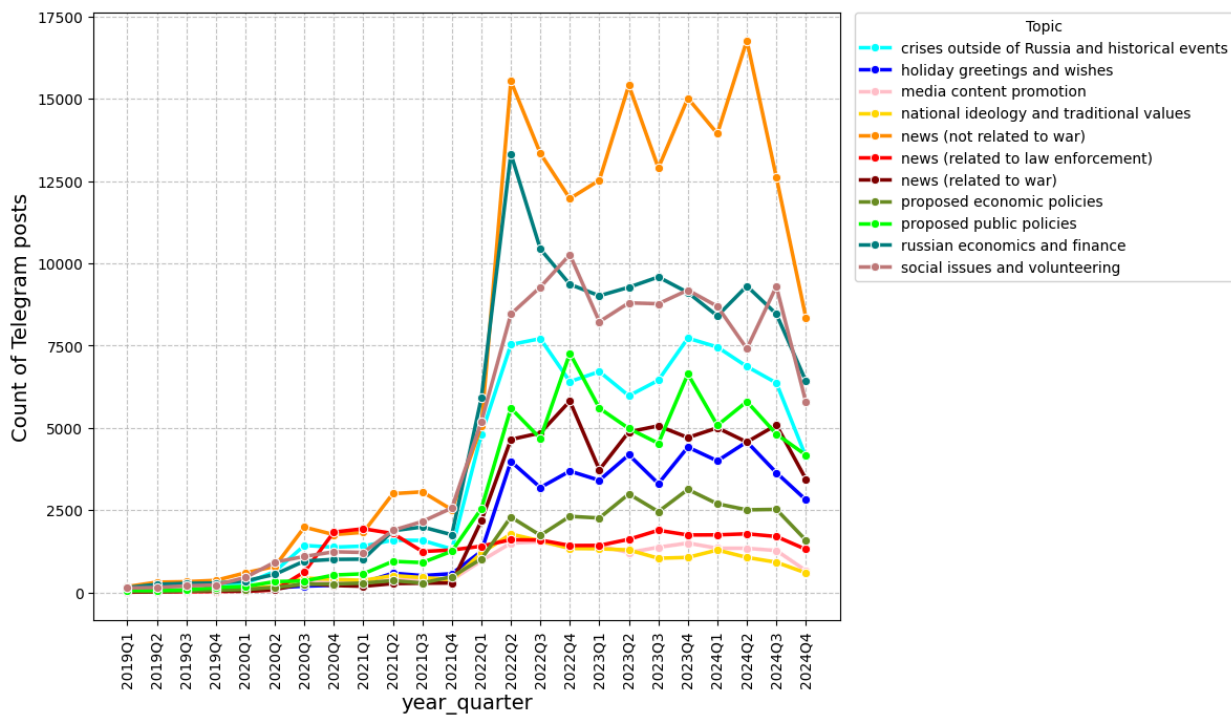

***Figure 5***. *Clustered topics for Russian policymakers.*

Together, these observations suggest that Russian policymakers focused on diverting public attention away from the invasion of Ukraine and the social and economic problems it caused. Their communication focused on unrelated news and positive developments, as well as foreign crises. More divisive subjects, including economic policies, related fines, and law enforcement actions, including those associated with anti-war protests, remained relatively marginal in the discourse. Although discussions about the war did occur after February 2022, they were not the main focus, and political communication concentrated on promoting messages about stability and unity within Russia, while highlighting turmoil elsewhere.

Figures 6 and 7 demonstrate that the communication of Ukrainian policymakers was rather different. Both in Ukrainian and Russian, the topic of the war-related news has been central since the beginning of the invasion. In the Russian subset, it dropped substantially already in spring 2022, only to peak again a year later, around the failed Ukrainian counteroffensive, and then declined sharply by the end of 2023. In the Ukrainian subset, the peak of communication about war-related news occurred at the end of 2022, following successful military operations, and then it decreased by the end of 2023. Such a dynamic can be an indicator of the tiredness of the war among both policymakers and society, and potentially, the pragmatic shift away from the war-related topics as the situation at the frontline continued to deteriorate. We observed a similar dynamic regarding another topic explicitly related to the war (i.e., humanitarian costs), which focused on the consequences of the Russian attacks against civilians and civilian infrastructure.

Another difference from communication in Russia was a strong focus of Ukrainian policymakers on international relations, especially in the Ukrainian subsample. Despite this topic being of particular importance for the country under attack, it was actively engaged with by policymakers commenting on international developments, from the elections in other countries to the international aid to Ukraine. The presence of this topic followed the similar dynamics as war-related news, highlighting that the two issues were usually discussed hand-in-hand. Besides, there was also more active communication regarding proposed economic policies, indicating that Ukrainian policymakers felt more entitled to comment on the key political issues (including cases when it involved criticising the authorities) and raise public awareness about them.

In terms of topical differences between the Ukrainian and Russian subsets, we observed a stronger presence of volunteering-related topics among Ukrainophone policymakers. By contrast, in the Russian subset (Figure 7), there was more communication regarding personal attacks and reflections on matters, from language-related issues to religion, indicating a particular communication style adopted by some Russophone policymakers active on Telegram (e.g., Maksym Buzhanskyi). These topics were pronounced before 2022, but after peaking around the invasion period, they dropped significantly, as war-related news dominated the discourse.

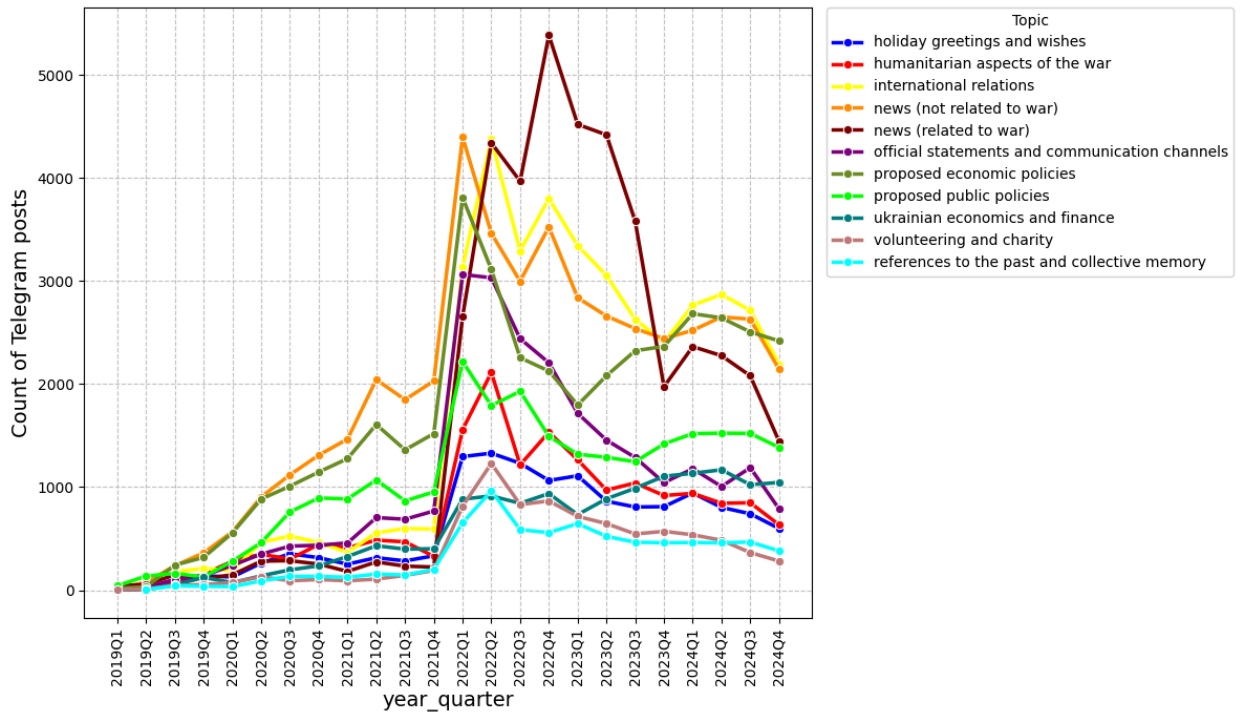

*Figure 6. Clustered topics for Ukrainian policymakers (posts in Ukrainian).*

Overall, we observed many similarities regarding topics communicated about by Ukrainian and Russian policymakers, as evidenced by a number of shared topics. However, there were major differences regarding the visibility of these topics, with Ukrainian policymakers putting the ongoing war at the center of their online discourse (at least until 2024), likely as part of public mobilization to resist the external aggression, whereas Russian policymakers omitted war and focused on the positive internal developments and negative views from outside Russia.

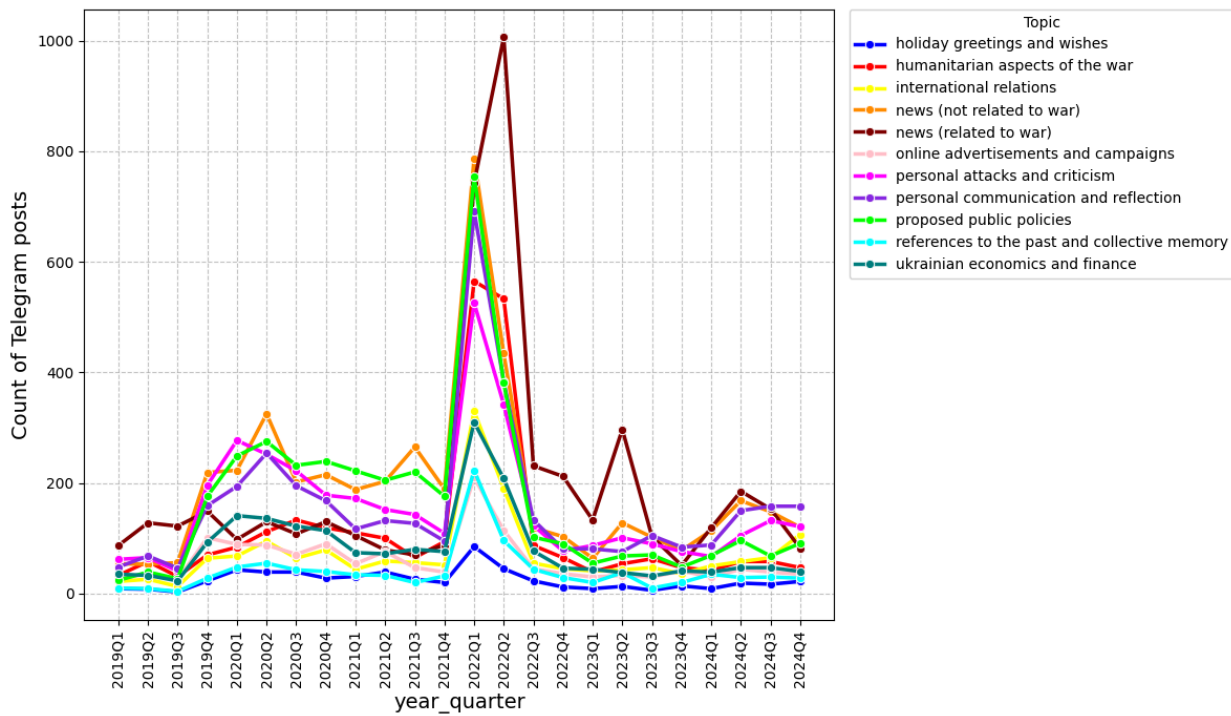

*Figure 7*. Clustered topics for Ukrainian policymakers (posts in Russian).

### *Actors and political communication on Telegram*

To answer RQ3, we first looked at the distribution of communication topics across policymaker groups. In the case of Russia (see Figure 8), we observed a rather similar distribution of topics among the largest parties (United Russia, Communist Party, LDPR, Just Russia, and, to a certain degree, New People): all four focused on topics not directly related to the war, such as general news, social care, economics, and foreign crises. Such consistency can be indicative of a consensus among major policymaker groups about what topics are desirable to communicate on.

The communication on the war was mostly driven by government-related Telegram accounts. Among the individual policymakers, it was most pronounced for Rodina, a marginal ultranationalist and pro-Kremlin party. Rodina was also an outlier in terms of more than half of its Telegram communication dealing with external crises and criticism of the West. Other smaller parties (e.g., Civic Platform and Party of Growth) put larger emphasis on policy-related subjects and economic matters, positioning themselves as more technocratic actors.

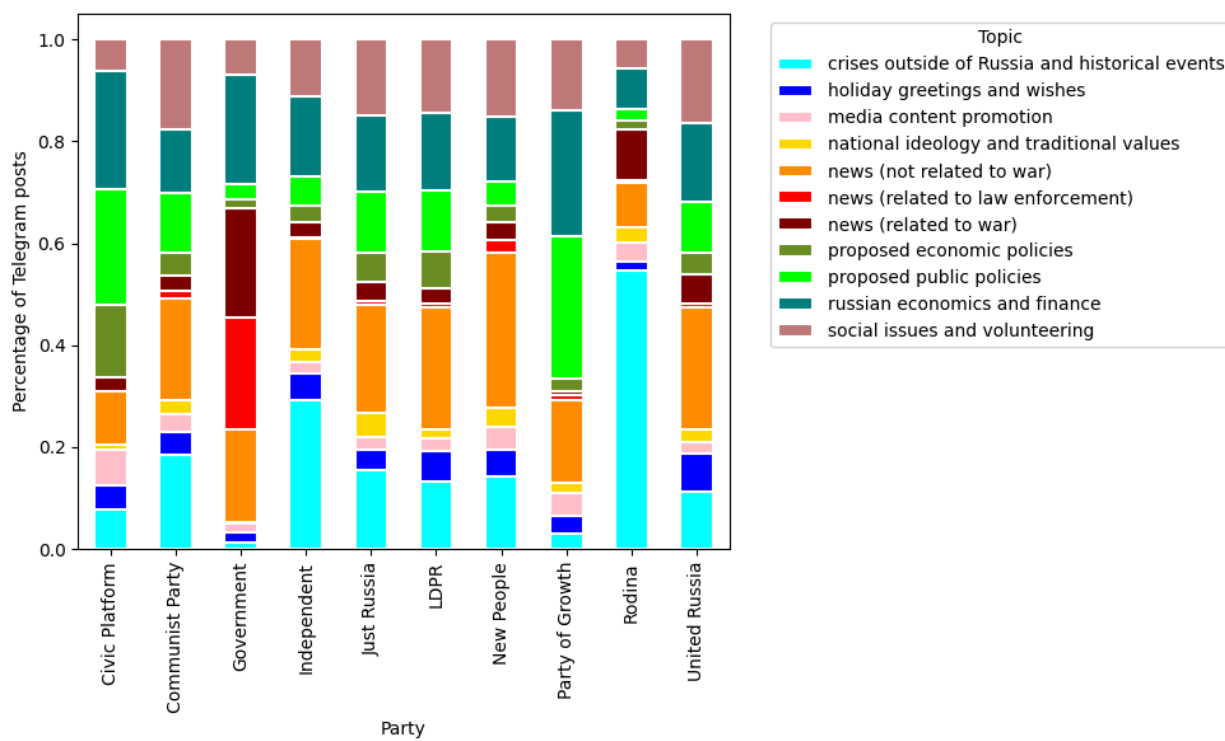

***Figure 8***. *Distribution of posts per topic per party for Russian policymakers.*

Among Ukrainian policymakers who communicated in Ukrainian (Figure 9), discussions about the war were more prevalent, especially among larger parties (i.e., Servant of the People, European Solidarity, Holos). Unlike Russia, communication on the war was not delegated to the governmental accounts. Overall, the differences between the government and major parties were minor, except that governmental accounts communicated more about the humanitarian costs of the war. The largest parties were also quite similar in terms of the large number of content dealing with international relations and news unrelated to the war. The international relations topic was especially prevalent for Batkivshina, a conservative and populist party led by former prime minister of Ukraine, Yulia Tymoshenko, constituting more than half of its Telegram content.

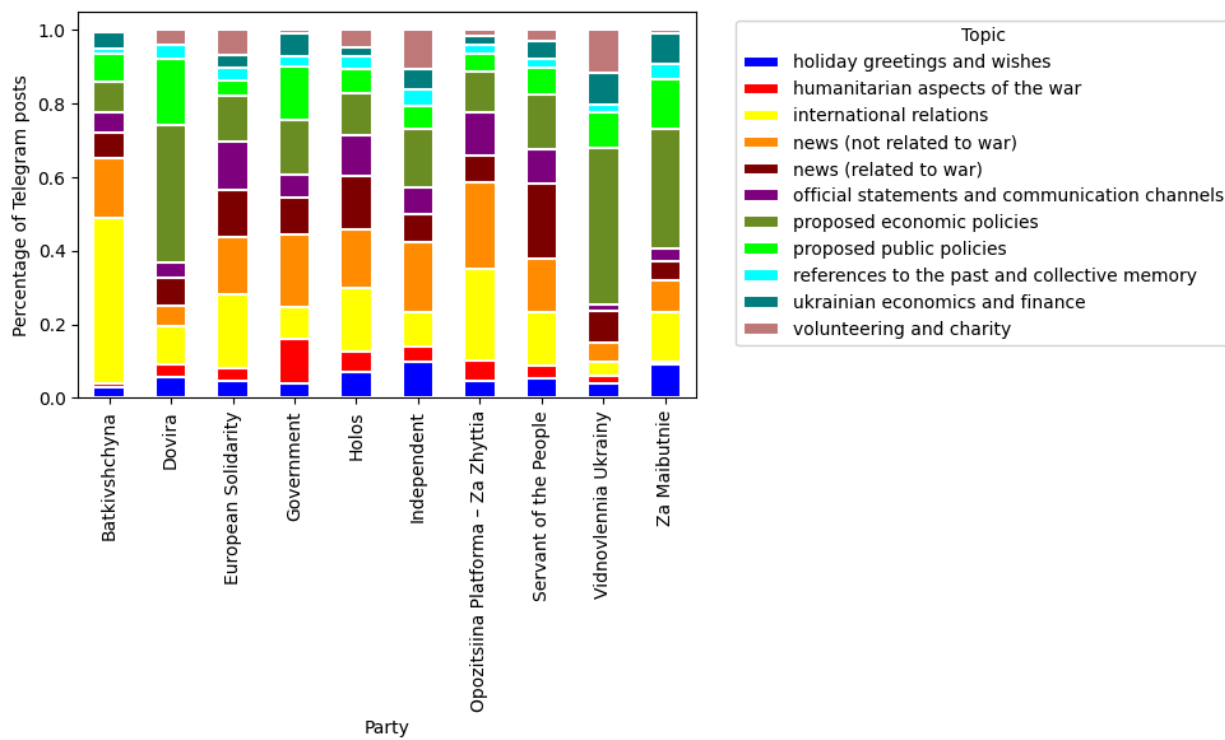

**Figure 9**. *Distribution of posts per topic per party for Ukrainian policymakers (Ukrainian language subsample).*

Smaller Ukrainian parties and parliamentary groups, such as Dovira, Vidnovlennia Ukrainy, and Za Maibutnie, focused on communicating about economic and public policies. This technocratic emphasis aligns with our observations for smaller Russian parties. For Ukraine, one reason for it could be these groups' connections to oligarchs and the pro-Russian party "Opozitsiina Platforma — Za Zhyttia." This background may explain why they prioritised economic policies over discussing current news. Interestingly, the distribution of topics communicated on by Russian-speaking Ukrainian policymakers from these parties was different (Figure 10). For Vidnovlennia Ukrainy and Za Maibutne, news regarding war and humanitarian aspects of the war were central subjects. For other parties, we observed a strong emphasis on public policies, news unrelated to war, and personal communication and reflection.

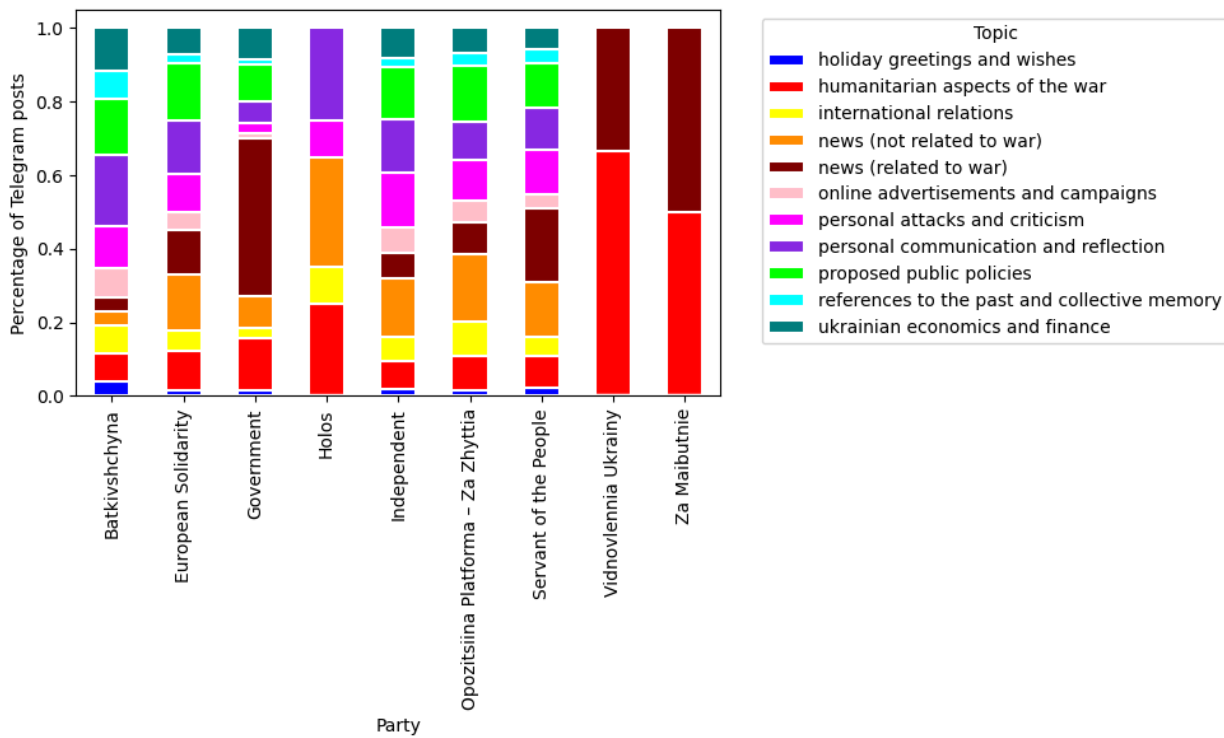

**Figure 10**. *Distribution of posts per topic per party for Ukrainian policymakers (Russian language subsample).*

After examining the communication at the party level, we then looked at how individual actors were involved in discussions on specific topics. In the case of Russian policymakers, the engagement varied strongly across different topics (Figure 11). For instance, communication about law enforcement, war, and Russian economics was led by the respective ministries. Additionally, governmental accounts actively communicated about social care and public policies. The institutional actors' focus on these specific topics suggests that they could be viewed as particularly important for the Russian state, with some of them being too sensitive for individual politicians to comment on extensively (e.g., law enforcement and war).

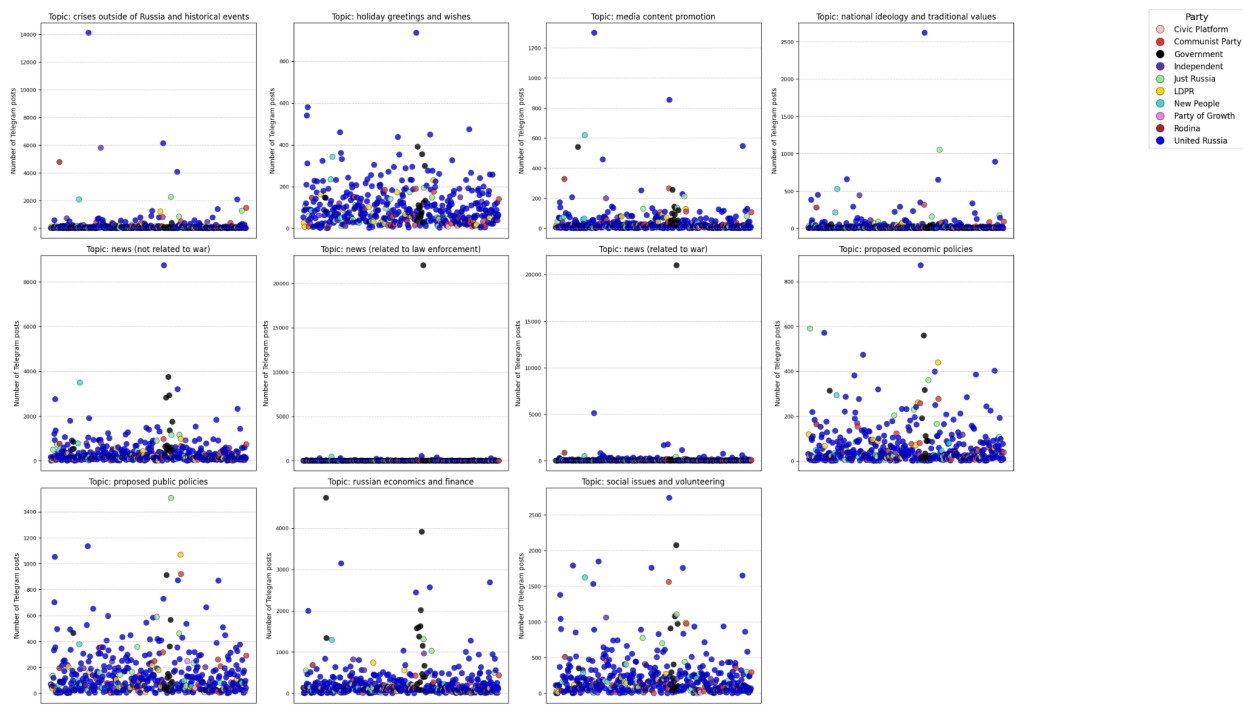

***Figure 11.*** *Count of posts per Russian policymaker per topic.*

In the case of other topics, such as national ideology, news unrelated to war, external crises, and media content promotion, communication was driven by a few individuals, particularly from the United Russia party. For the former two topics, the communication was driven by Oleg Matveychev, a Russian blogger and policymaker known for his provocative claims (e.g., calling for the return of Russian colonies or raping of relatives of protesters in Kazakhstan). For the latter two topics, the most active actor was Vyacheslav Nikonov, a Russian political scientist and policymaker, known for declaring the Russian invasion of Ukraine a holy war. Other policymakers actively included in the communication on these topics were Nikolai Valuev, a former professional boxer, Yevgeny Popov, a Russian journalist and propagandist, and Roza Chemeris, a member of the New People party. Such a composition, especially considering the strong presence of media provocateurs like Matveychev, Popov, and Nikonov, suggests that these particular topics served as a distraction of public attention during the war.

Finally, we observed several topics for which communication involved a broader range of policymakers. Examples included proposed public and economic policies, social care, and holiday greetings. These topics can be viewed as less sensitive and, thus, more policymakers could contribute to them. Interestingly, for public policies, the most active poster was Sergey Mironov, the leader of the Just Russia party. It was the only topic where the most active poster was not from the United Russia or one of the governmental accounts.

For the Ukrainian policymakers posting in Ukrainian (see Figure 12), we observe certain similarities with the Russian case. There are topics for which communication is driven by governmental actors, specifically the Ministry of Internal Affairs (e.g., humanitarian aspects of the war, news unrelated to war, and references to the past and collective memory), the Ministry of Education and Science (proposed public policies), the Ministry of Finance (Ukrainian economics and finance), and the Ministry of Youth and Sport (news unrelated to war). One important distinction, however, is a prominent role of policymakers outside the ruling party in political communication. For instance, Oleksii Honcharenko, a member of European Solidarity, was among the most active users on the topics of news (both related and unrelated to war), humanitarian aspects, and references to the past. Other members of European Solidarity, such as Petro Poroshenko, Sofia Fedyna, and Iryna Herashchenko, were also among the key communicators on references to the past, economic policies, volunteering, and holiday greetings.

In the case of Ukrainian policymakers posting in Russian (see Figure 13), we observe a tendency for individual actors to drive communication on specific topics. Most of these power users were from the Servant of the People party or independent policymakers, with a few active posters from European Solidarity. For all topics but the public policies, online advertisement, and war-related news, the top poster was Maksym Buzhanskyi, a Servant of the People member and a blogger. For the former two topics, the discussion was driven by Oleksandr Dubinsky, a former member of the Servant of the People, who was expelled from the party in 2021 (and, hence, marked as an independent) and arrested for treason in 2023.

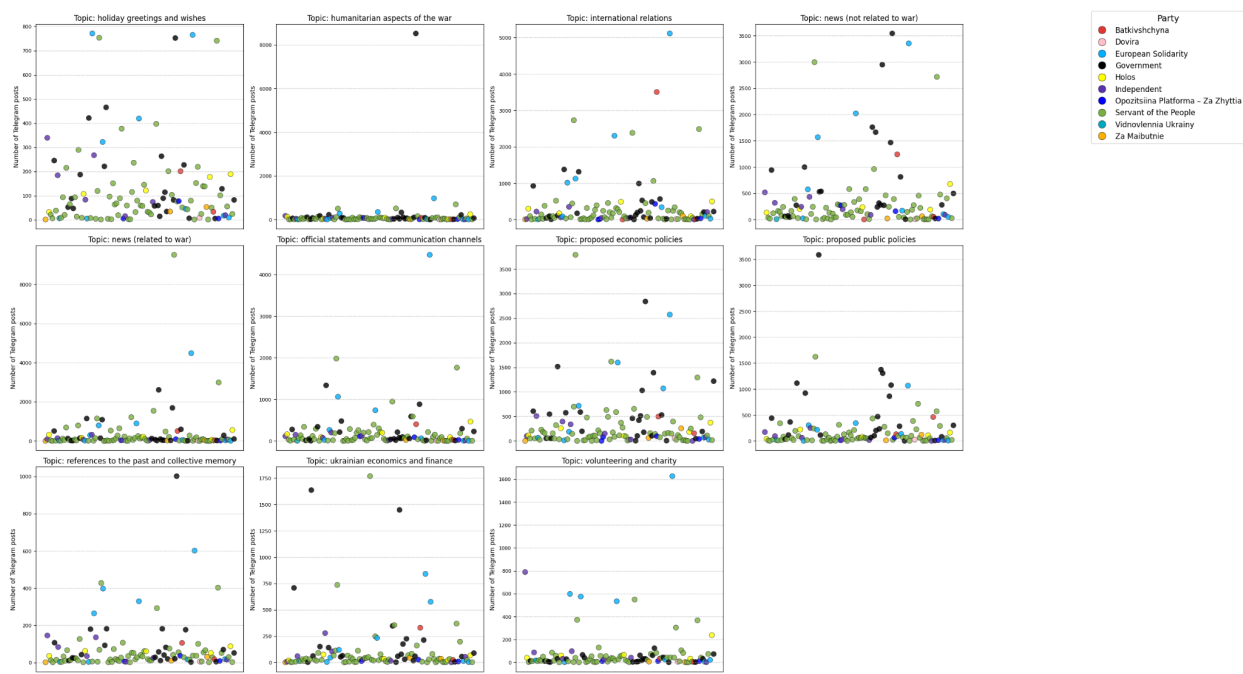

***Figure 12.*** *Count of posts per Ukrainian policymaker per topic (Ukrainian language subsample).*

For the war-related news, the top poster was Yuri Misyagin, another parliamentarian from the Servant of the People party and an active supporter of the Ukrainian armed forces since 2014. Other policymakers actively engaged in Russophone communication on Telegram were Oleksii Honcharenko (European Solidarity) and Viktor Chornyi (Servant of the People). Finally, a few policymakers were active only in relation to a specific topic, such as proposed public policies (Danilo Hetmantsev, Servant of the People) or volunteering (Viktoriya Hryb, independent).

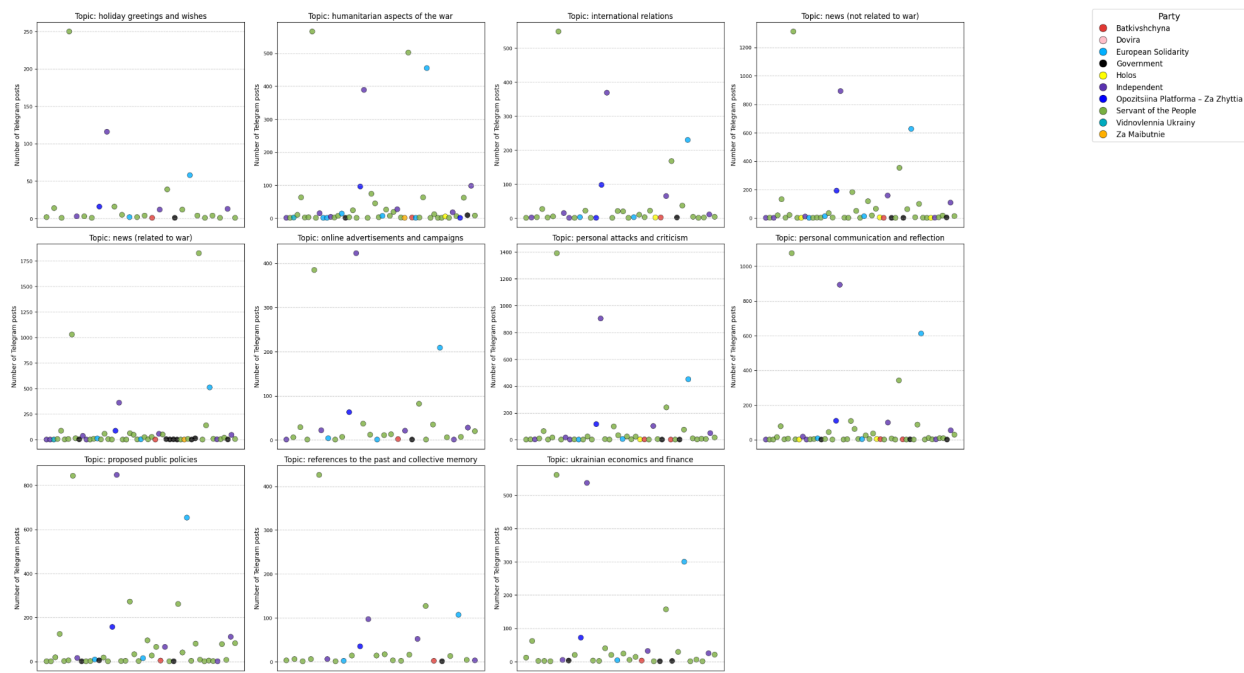

***Figure 13.*** *Count of posts per Ukrainian policymaker per topic (Russian language subsample).*

## Discussion

In this article, we looked at how political communication on Telegram can shape perception and legitimization of wars in countries with high internet penetration, where social media and messengers serve as a crucial means of communication between the elites and the public. Using a dataset of Telegram posts from Ukrainian and Russian policymakers before and after the Russian invasion in February 2022, we examined how political communication changes in both the aggressor and the defending countries at the time of war.

Our first finding is that the full-scale invasion resulted in a major spike in activity on Telegram. Both in Ukraine and Russia, it caused a growth in the absolute number of posts compared with the period before the invasion and the number of policymakers posting on Telegram. Such a growth was particularly pronounced for institutions and policymakers from the ruling party, especially in Russia. This increase suggests that the invasion has been viewed as a major challenge for communication in both countries, prompting policymakers to respond.

Our second finding regards the differences between Ukrainian and Russian policymakers in their communication topics. Among Ukrainian policymakers, the discussion of the war featured prominently since the beginning of the invasion, together with other news and international relations. The discussions of the war peaked in 2022 and has declined since to the levels of other prominent topics; such a change in communication can be attributed to multiple factors, from the normalization of the war to the growing tiredness of the population from the war. In contrast, Russian policymakers adopted a different strategy, with elites largely avoiding the topic of the war and instead focusing on unrelated news, particularly emphasizing crises in the West. The core aim of this strategy was likely to divert public attention away from the war and its toll. This communication remained consistently applied throughout the period of observation.

Our third finding regards the differentiated involvement of policymakers in wartime communication. On the level of parties, we identified differences between large parties, which engaged with a broad range of topics, and smaller parties, which specialized in a few issues (e.g., economy or traditional values). On the level of individuals, we observed differences depending on the topic: Some topics were dominated by small groups of (predominantly male) power users, whereas for other topics the engagement was more dispersed.

Together, these findings provide important insights into how elite-driven wartime communication functions online. Specifically, they showcase that autocracies, like Russia, may have an easier time adapting and streamlining governmental communication to the state of emergency compared to democracies, even in less top-down and controlled socially mediated environments. Besides, we find that while war-related news constitutes an important part of wartime communication, such communication is not limited to violence-related discussions. Both in Ukraine and Russia, policymakers kept communicating on other topics, from the economy and public policies to holidays greetings, thus interweaving the war discourse with other issues.

It is important to note several limitations of this study. We focus on the platform that until now has had little moderation and is not viewed as mainstream as some other platforms (e.g., Facebook). However, the developments since Trump's second term demonstrate how niche platforms (e.g., Truth Social) can be used by elites to reach large audiences. Furthermore, considering the proportion of policymakers using Telegram in Ukraine and Russia, the platform can hardly be viewed as non-mainstream in these two countries. Another limitation is that we looked only at policymakers, whereas there are many other actors involved in political communication (e.g., activists, propagandists, or military bloggers). While these actors are important, we argue that they have less capacity to translate discourse into policies (especially in Russia), so policymakers still play a key role in deciding which interpretations of the war will be picked up by other media and shape societal expectations about the war and its outcomes.

There are also several directions for future research, which we believe are important to note. One of them is investigating coordinated sharing behavior among Ukrainian and Russian policymakers to identify networks of highly coordinated political actors and the content they amplified (Righetti & Balluff, 2025). These could include not only textual information but also images and videos shared strategically by policymakers. Another direction is conducting time series analyses (e.g., using Granger-causality tests) to examine the temporal dynamics of selected topics on Telegram and their connection to offline developments.

Notwithstanding these limitations, our study makes a valuable contribution to a scholarly discussion about elite-driven wartime political communication online. By focusing on policymakers in Ukraine and Russia, we provide a much-needed comparative (and longitudinal) perspective on the evolution of political discourse before and after the invasion. Furthermore, we introduce a meticulously curated dataset containing all Telegram channels of current policymakers in both countries, which can be a valuable resource for future studies on the topic.

**Online appendix**

**Appendix 1**: **Implementation of the BERTopic for topic modelling**

In this appendix, we describe how we implemented BERTopic to process Telegram content produced by Ukrainian and Russian policymakers from 2019 to 2024. Specifically, we describe how we implemented document embeddings and reduced their dimensionality, clustered the embeddings, vectorized documents, identified the most representative terms, selected topics, and reduced outliers. By combining all these steps, we aimed to improve the performance of the BERTopic for topic identification.

As a first step, we used document embeddings, which we generated using the pre-trained multilingual sentence transformer model, *paraphrase-multilingual-MiniLM-L12-v2* (Reimers & Gurevych, 2019). We selected it due to the general effectiveness of the sentence transformers; the choice of the model was due to it being particularly fitting for our case because we wanted to compare topics emerging from the datasets in two different languages. To reduce the dimensionality of embeddings, we used the Uniform Manifold Approximation and Projection technique (McInnes et al., 2018). Following the standard practices, we reduced the dimensionality of the embeddings to 5 components, while setting the number of neighbors to 15. To mitigate potential overclustering, we set the minimum distance parameter to 0.1. The cosine metric was used to measure the distance between embeddings.

To cluster the reduced embeddings, we applied the High-Definition Density-Based Spatial Clustering of Applications with the Noise technique (Campello et al., 2013). We opted for a dynamically determined minimal cluster size (i.e., based on the document dataset size) to enable a more consistent cluster granularity across different datasets. Specifically, we set it to 0.01% of the total number of documents in each dataset. This resulted in a minimum cluster size of approximately 80 documents for the larger Russian dataset (consisting of 800,000+ documents after the filtering) and 6 documents for the smaller Russian subset of the Ukrainian dataset (approximately 60,000 documents after filtering). We applied the Euclidean metric for distance calculation and the excess of mass method for cluster selection. Finally, we enabled the prediction of cluster membership for new data points to identify the most stable clusters.

Prior to topic extraction, we vectorized documents in our datasets using CountVectorizer. For each three datasets — the Russian dataset, the Russian subset of the Ukrainian dataset, and the Ukrainian subset of the Ukrainian dataset — we removed stop words. For Russian datasets, we used the list of Russian stopwords from the Python *NLTK* library, whereas for the Ukrainian dataset, we applied a list compiled by Serhii Kupriienko (2021). We set the minimum document frequency to 0.01% of the total number of documents, mirroring the logic for determining the minimal cluster size and ensuring that only terms appearing in a sufficient number of documents are considered. To capture frequent word combinations, we included both unigrams and bigrams.

To identify the most representative terms for each cluster, we employed the Class-based TF-IDF (c-TF-IDF) approach implemented in BERTopic. Recognizing potential limitations in the provided stop word lists for Ukrainian and Russian, we further down-weighted highly frequent words within the clusters that might not be effectively captured by standard stop word lists, as suggested by Grootendorst (2022). Finally, we allowed BERTopic to automatically determine the optimal number of topics based on the clustering results using the "auto" parameter and reduced the outliers using the c-TF-IDF technique. It resulted in a large number of topics: 205 for the Russian subset of the Ukrainian dataset, 483 for the Ukrainian subset of the Ukrainian dataset, and 399 for the Russian dataset. To further decrease the number of topics and make the results of the topic modelling more interpretable, two authors who are experts on Russian and Ukrainian politics manually grouped topics into larger clusters. While doing so, they discussed which topics can be merged to which clusters and iteratively examined each other's clustering to minimize the effect of individual bias.

**Appendix 2**: **Supplementary information about Ukrainian and Russian policymakers**

In this appendix, we provide a short overview of the political systems in Ukraine and Russia, together with a brief description of the political parties to help contextualize our analysis.

*Political system in Ukraine*

Ukraine is a semi-presidential republic with a multi-party political system. The government is divided into the executive, legislative, and judicial branches. The President serves as the head of state and is responsible for foreign policy and national security. The head of government is the Prime Minister, who leads the Cabinet of Ministers that serves as an executive power. Finally, the unicameral parliament, the Verkhovna Rada, is composed of 450 members and serves as a legislative power responsible for adopting laws, approving the state budget, and overseeing the government's work.

The latest presidential and parliamentary elections in Ukraine happened in 2019 and resulted in a landslide victory for Volodymyr Zelensky and his party, the Servant of the People. While the next elections should have happened in 2023 (for Verkhovna Rada) and 2024 (for the Peesident), due to the martial law declared following the full-scale invasion by Russia, all elections have been postponed. As of 2024, the following nine political parties and parliamentary groups are represented in the Ukrainian parliament:

*Servant of the People*: This is a populist and centrist pro-European party led by Ukrainian president, Volodymyr Zelenskyy. It is the ruling party that received the majority in the Parliament, winning 254 seats during the 2019 election. Its programme focuses on digitalization, European integration, and anti-corruption measures. Initially, it also pursued the programme of national unity, claiming that its aim is to unite Ukrainians and counter the divides between different parts of the country.

*European Solidarity*: This party is led by the Ukrainian former president, Petro Poroshenko, and its programme is based on liberal conservatism and civic nationalism. It acquired 25 seats during the 2019 elections. The party advocates for Ukraine's membership in NATO and the European Union, a strong military, and a free-market economy. Unlike the Servant of the People, which had strong support across different parts of Ukraine, the European Solidarity has stronger support in the Western regions.

*Batkivshchyna (Motherland)*: Led by former Prime Minister Yulia Tymoshenko, this is a centre-left populist party that advocates for social welfare, national solidarity, and Christian values. It received 26 seats in the parliament after the 2019 elections. The party's socioeconomic programme puts a strong emphasis on social justice, including the need to revise the existing pension system, and energy independence for Ukraine.

*Opozitsiina Platforma – Za Zhyttia (Opposition Platform - For Life)*: It is a pro-Russian, social-democratic, and Eurosceptic party that focuses in its programme on the protection of the Russophone population of Ukraine. In the 2019 election, it won 43 seats, becoming the largest opposition party. The party was later suspended and then banned in 2022 due to the war with Russia, and its former members have since formed new parliamentary groups.

*Holos (Voice)*: Founded in 2019 by musician Svyatoslav Vakarchuk, it is a center-right pro-European party with a strong anti-corruption stance. It won 20 seats in the 2019 election. The party supports the establishment of a land market, the privatization of the industrial enterprises, and actively campaigns against illegal customs schemes.

*Za Maibutnie (For Future)*: This parliamentary group was established in 2019, and in 2020, it became a political party on the basis of the pre-existing "Ukraine of Future" party. It has 17 seats in the parliament and is associated with the oligarch Ihor Kolomoyskyi. The group's platform is centered on economic nationalism and regional self-governance.

*Vidnovlennia Ukrainy (Restoration of Ukraine)*: This parliamentary group was formed in the Verkhovna Rada in 2022, primarily from the former members of the banned "*Opozitsiina Platforma – Za Zhyttia*". It consists of 17 members and has an obscure ideological platform that primarily focuses on the (economic) rebuilding of Ukraine after the war.

*Dovira (Trust)*: This parliamentary group was established at the end of 2019. It consists of 19 non-party deputies elected via single-member constituencies. Similar to other parliamentary groups, it has a rather abstract political programme that advocates for a professional and expertise-based approach to legislation and support of regions.

*Independent*: Members of the parliament who are not affiliated with any official political party or parliamentary group. Some of them were elected as self-nominees or left their former parties.

### Political system in Russia

Russia's political system is a federal semi-presidential republic with a strong executive branch. The President, as head of state, holds a dominant position, wielding significant authority to appoint the Prime Minister and cabinet, influence legislation, and serve as commander-in-chief. The legislative body, the Federal Assembly, is bicameral, comprising the State Duma (the lower house; 450 members) and the Federation Council (the upper house; 170 members). The Federation Council members are not elected but assigned by Russia's federal subjects. While Russia is formally a multi-party system, the political landscape is largely controlled by the pro-government party, United Russia, with limited independence of systemic "opposition" (and close to no non-systemic opposition due to state repressions) and a highly centralized government structure.

The most recent presidential election in Russia happened in the spring of 2024, while the last parliamentary election was in the autumn of 2021. The next presidential election is scheduled for 2030, and the next parliamentary election is expected in 2026.

*United Russia*: Established in 2001, through a merger of several parties, United Russia is a ruling pro-state party. It has a majority in the State Duma with 324 seats after the 2021 elections. The party's centrist programme promotes a pro-presidential agenda, emphasizing political stability, national sovereignty, and conservative values. Its core focus is on supporting the policies of the current government and Vladimir Putin, including Putin's foreign policy.

*Communist Party*: This left party was established in 1993 as the successor to the Soviet Union's Communist Party. It is the second-largest party in the State Duma that won 57 seats during the last elections. The party's programme declares its aim as the restoration of a socialist state that involves the nationalization of industry, increased social welfare, and opposition to capitalism.

*LDPR*: This right-wing populist party was established in 1992 and was centered around its leader, Vladimir Zhirinovsky, until his passing in 2022. After the 2021 elections, it has 21 seats in the State Duma. The party's programme is based on ultranationalist and populist ideas, arguing for the revival of Russia as a great power, strengthening of the central government, and protection of ethnic Russians outside Russia.

*A Just Russia – For Truth*: This centre-left party was established in 2006, from a merger of several conservative Russian parties, and more mergers to follow (e.g., the 2021 merger with the Patriots of Russia and For Truth parties). In the 2021 elections, it won 27 seats. The party's programme combines social conservatism and left populism, advocating for the welfare state, intensification of anti-corruption campaigns, and restoration of Russia's status as a great power (and opposition to the West).

*New People*: This centre-right party was established in 2020, presumably as a spoiler party to attract protest voters. It received 13 seats in the 2021 elections. The party's programme focuses on liberalization of state policies (particularly, in the economic sphere) and limiting the power of the state apparatus. It claims to represent the interests of self-employed citizens and small businesses and opposes certain policies of the government (e.g., regarding the increase of the state control over the public sphere).

*Civic Platform*: The party has been established in 2012 by businessman Mikhail Prokhorov, and has one seat in the State Duma. This is a conservative party focused on supporting small and medium-sized businesses and promoting liberal-conservative values. It advocates for economic liberalization and greater transparency in government, often focusing on local and regional elections.

*Rodina (Motherland)*: Established in 2003 as a bloc of predominantly right-leaning and conservative groups, Rodina was disbanded and then re-established in 2012. After the 2019 elections, it has one seat in the State Duma. The programme of this far-right populist and ultranationalist party focuses on advocating for a strong, centralized state and promoting a conservative nationalist ideology.

*Party of Growth*: Established in 2016 on the basis of the Right Cause party, this liberal-conservative party merged with the New People party in 2024. Before its merger, it had one seat in the State Duma. The party positioned itself as a pro-business party and advocated for free market reforms and privatization. Despite the party occasionally criticising governmental policies, its head, Boris Titov, serves as a presidential commissioner for entrepreneurs' rights.

*Independent*: Similar to Ukraine, there are policymakers in the Russian parlimated who are not affiliated with a party or a parliamentary group. Currently, there are five independent parlamentarians in the State Duman.